\documentclass[review]{elsarticle}

\usepackage{lineno, hyperref}
\modulolinenumbers[5]
\usepackage{amsmath}
\usepackage{siunitx}
\usepackage{amssymb}
\usepackage{booktabs}
\usepackage{multirow}
\usepackage{subcaption}
\usepackage{color, soul}

\journal{NIM A}

\begin{document}

\begin{frontmatter}

\title{Embedding the Timepix4 in Micro-Pattern Gaseous Detectors}

\author[cern]{L. Scharenberg\corref{cor1}}
\ead{lucian.scharenberg@cern.ch}
\cortext[cor1]{Corresponding author}

\author[cern]{J. Alozy}
\author[cern]{W. Billereau}
\author[cern]{F. Brunbauer}
\author[cern]{M. Campbell}
\author[cern]{P. Carbonez}
\author[cern,hiskp]{K.J. Fl\"{o}thner}
\author[hip]{F. Garcia}
\author[cern,chuv]{A. Garcia-Tejedor}
\author[cern,chuv]{T. Genetay}
\author[nikhef]{K. Heijhoff}
\author[cern]{D. Janssens}
\author[cern]{S. Kaufmann}
\author[cern]{M. Lisowska}
\author[cern]{X. Llopart}
\author[cern]{M. Mager}
\author[cern]{B. Mehl}
\author[cern,bonn]{H. Muller}
\author[cern]{R. de Oliveira}
\author[cern]{E. Oliveri}
\author[cern,erlangen]{G. Orlandini}
\author[ess,cern]{D. Pfeiffer}
\author[cern]{F. Piernas Diaz}
\author[cern]{A. Rodrigues}
\author[cern]{L. Ropelewski}
\author[ess]{J. Samarati}
\author[nikhef]{M. van Beuzekom}
\author[cern]{M. van Stenis}
\author[cern]{R. Veenhof}
\author[unige]{and M. Vicente}

\address[cern]{European Organization for Nuclear Research (CERN), 1211 Geneva 23, Switzerland}
\address[hiskp]{Helmholtz-Institut f\"{u}r Strahlen- und Kernphysik, University of Bonn, Nu\ss{}allee 14-16, 53115 Bonn, Germany}
\address[hip]{Helsinki Institute of Physics, P.O. Box 64, FI-00014 University of Helsinki, Finland}
\address[chuv]{Institute of Radiation Physics, Centre hospitalier universitaire vaudois (CHUV), Rue du Grand-Pr\'{e} 1, 1007 Lausanne, Switzerland}
\address[nikhef]{Nikhef, Science Park 105, 1098 XG Amsterdam, The Netherlands}
\address[bonn]{Physikalisches Institut, University of Bonn, Nu{\ss}allee 12, 53115 Bonn, Germany}
\address[erlangen]{Friedrich-Alexander-Universit{\"a}t Erlangen-N{\"u}rnberg, Schlo\ss{}platz 4, 91054 Erlangen, Germany}
\address[ess]{European Spallation Source ERIC (ESS), Box 176, SE-221 00 Lund, Sweden}
\address[unige]{D\'{e}partement de physique nucl\'{e}aire et corpusculaire, University of Geneva, 24 quai Ernest-Ansermet, 1205 Gen\`{e}ve 4, Switzerland}

\begin{abstract}
The combination of Micro-Pattern Gaseous Detectors (MPGDs) and pixel charge readout enables specific experimental opportunities.
Using the Timepix4 for the readout is advantageous because of its size (around $\SI{7}{cm^2}$ active area) and its Through Silicon Vias.
The latter enables to connect to the Timepix4 from the back side.
Thus, it can be tiled on four sides, allowing it to cover large areas without loss of active area.

Here, the first results of reading out MPGDs with the Timepix4 are presented.
Measurements with a Gas Electron Multiplier (GEM) detector show that event selection based on geometrical parameters of the interaction is possible, X-ray imaging studies can be performed, as well as energy and time-resolved measurements.
In parallel, the embedding of a Timepix4 into a micro-resistive Well (\textmu{}RWell) amplification structure is explored.
The first mechanical tests have been successful.
The status of the electrical functionality is presented, as well as simulation studies on the signal induction in such a device.
\end{abstract}


\end{frontmatter}


\section{Introduction}
Combining gaseous detectors with high-granularity charge readout offers very specific possibilities, which otherwise could not be achieved.
Examples are high-resolution tracking of low-momentum particle beams (i.e. requiring low-material budget), X-ray polarimetry, especially in the low-energetic photon regime ($E_\text{photon} < \SI{2}{keV}$), as well as rare-event searches that profit from event-selection based on geometrical parameters.

For the readout, hybrid pixel Application-Specific Integrated Circuits (ASICs) are used, here specifically the Timepix4 \cite{timepix4}.
The goal is to embed the Timepix4 into a gaseous amplification stage using standard micro-pattern technologies for Printed Circuit Board (PCB) manufacturing.
Based on an idea presented at the Micro-Pattern Gaseous Detectors conference in 2022 \cite{embedding}, the concept has been explored in more detail, leading to a new research line within Work Package 2 of CERN's EP R\&D Programme \cite{ep-rnd-2}.
While the main objective is to investigate the embedding technology (Section~\ref{sec:embedding}), also other studies have been conducted with the amplification stage being separated from the readout stage (Section~\ref{sec:gempix} on the GEMPix4).

\section{Embedding approach}
\label{sec:embedding}

Following an introduction to the Timepix4, the embedding approach, its current status, as well as simulations on the signal induction are presented in this section.

\subsection{Timepix4}

The Timepix4's pixel matrix consists out of $\num{448}\times\num{512}$ square pixels with $\SI{55}{\micro m}$ pitch.
This active area is $\num{24.6}\times\SI{28.2}{mm^2} \approx \SI{7}{cm^2}$.
Each pixel comprises a Charge-Sensitive Amplifier (CSA) followed by a discriminator and a digitisation-logic.
For the use with MPGDs, no semiconductor sensor is bump-bonded to the pixel pads.
Instead, they are used as charge collection pads within the amplification structures.

The Timepix4 can be read out in two modes.
In the frame-based mode, each pixel counts the number of signals crossing the discriminator's threshold within a pre-defined amount of time.
In the data-driven mode, the induced signal charge is provided as Time-Over-Threshold (TOT) with $\approx\num{700}$ electron resolution (FWHM), simultaneously to the signal's Time-Of-Arrival (TOA) recorded with $\SI{200}{ps}$ time bins.
In addition, the Timepix4 contains the possibility of using Through Silicon Vias (TSVs).
While the Timepix4 is typically read out using wire bonds, the TSVs allow it to connect to the Timepix4 from the back side.
This is essential and the Timepix4's most crucial feature for the embedding, which would be mechanically impossible with wire bonds.
For reading out and controlling the Timepix4, the Speedy Pixel Detector Readout Version-4 (SPIDR4) \cite{spidr,spidr4}, developed at Nikhef, was used.

\subsection{Embedding concept}

\begin{figure}[t!]
    \centering
    \includegraphics[width = \columnwidth]{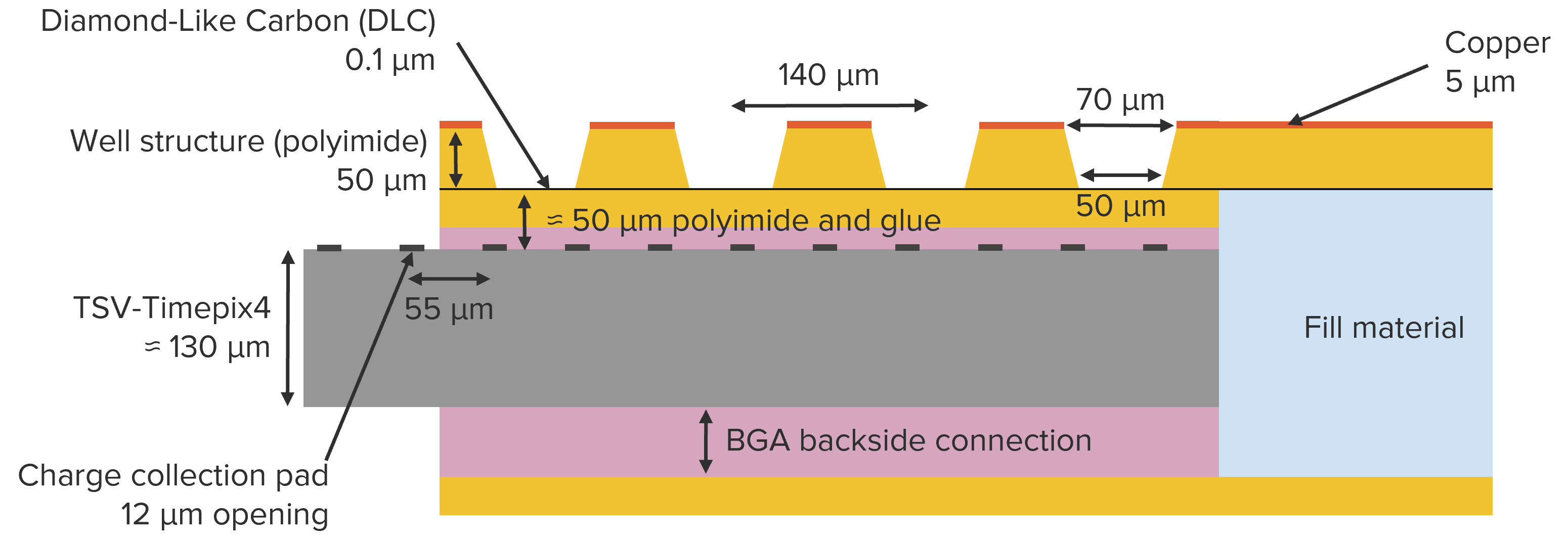}
    \caption{Schematic illustration of a Timepix4 embedded within two polyimide foils, one of them being the \textmu{}RWell amplification stage and the other one the PCB to connect to the readout system.}
    \label{fig:embedding-concept}
\end{figure}

Integrating a micro-pattern gaseous amplification stage to a pixel ASIC, specifically the Timepix \cite{timepix} has been successfully demonstrated before \cite{pixel-tpc}, however, requires wafer post-processing \cite{gridpix}.
The use of standard PCB technologies follows the approach of the  `MAPS foils' \cite{maps-foil}, where Monolithic Active Pixel Sensors (MAPS) have been laminated within two polyimide foils.
However, instead of laminating the Timepix4 within two `standard' polyimide foils, one of them can be replaced with a micro-resistive Well (\textmu{}RWell) \cite{urwell} (Fig.~\ref{fig:embedding-concept}).
Being itself polyimide-based, the \textmu{}RWell technology ensures the embedding process.
In addition, it provides discharge protection to the detector through the resistive layer of Diamond-Like Carbon (DLC), as well as the electronics through the insulation layer between the DLC and the Timepix4.

\subsection{Production status}

\begin{figure}[t!]
    \centering
    \begin{subfigure}{0.45\columnwidth}
        \centering
        \includegraphics[width = \columnwidth]{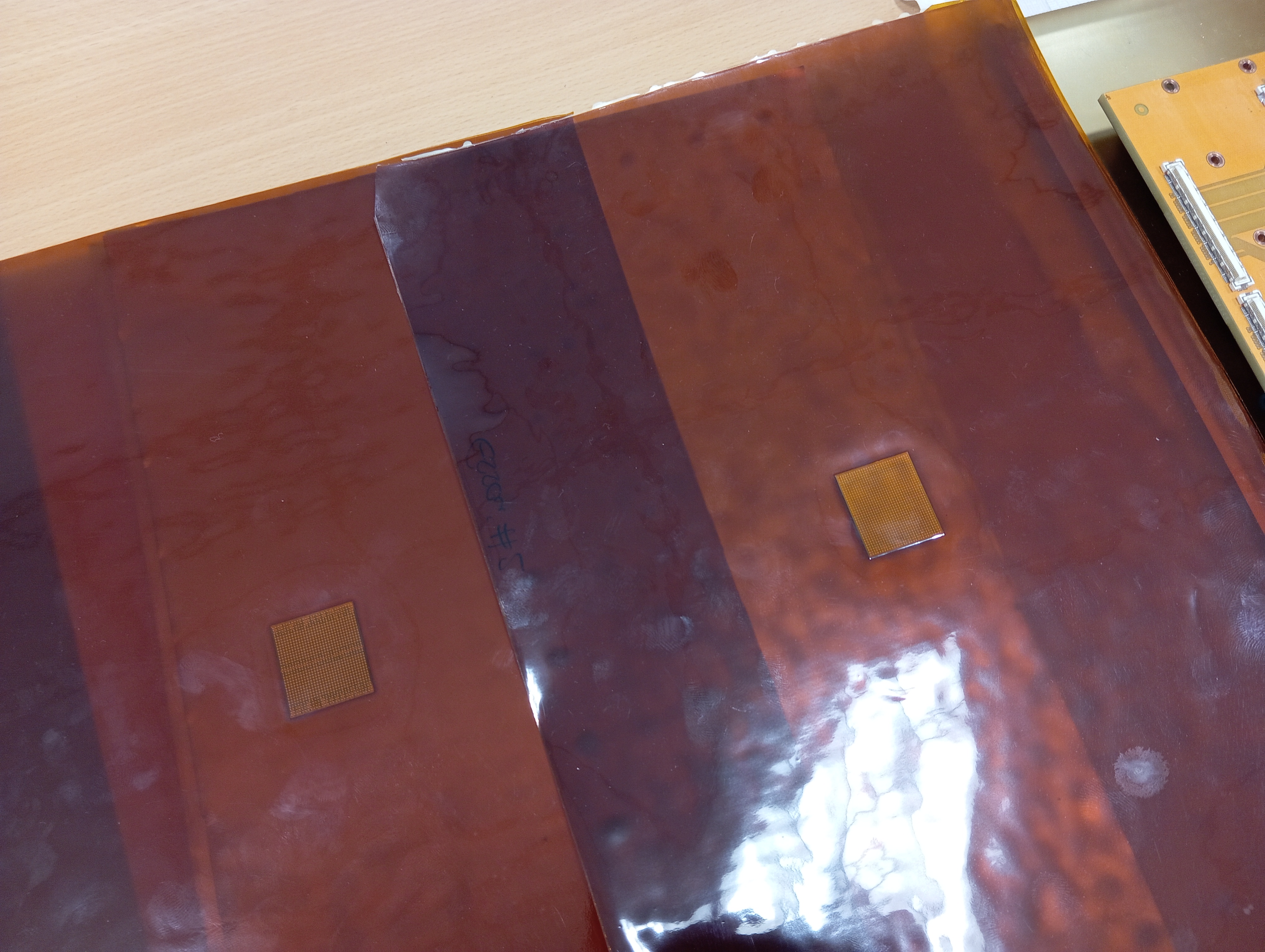}
        \caption{Mechanical embedding}
        \label{fig:embedding-steps-0}
    \end{subfigure}
    \hspace{0.05\columnwidth}
    \begin{subfigure}{0.45\columnwidth}
        \centering
        \includegraphics[width = \columnwidth]{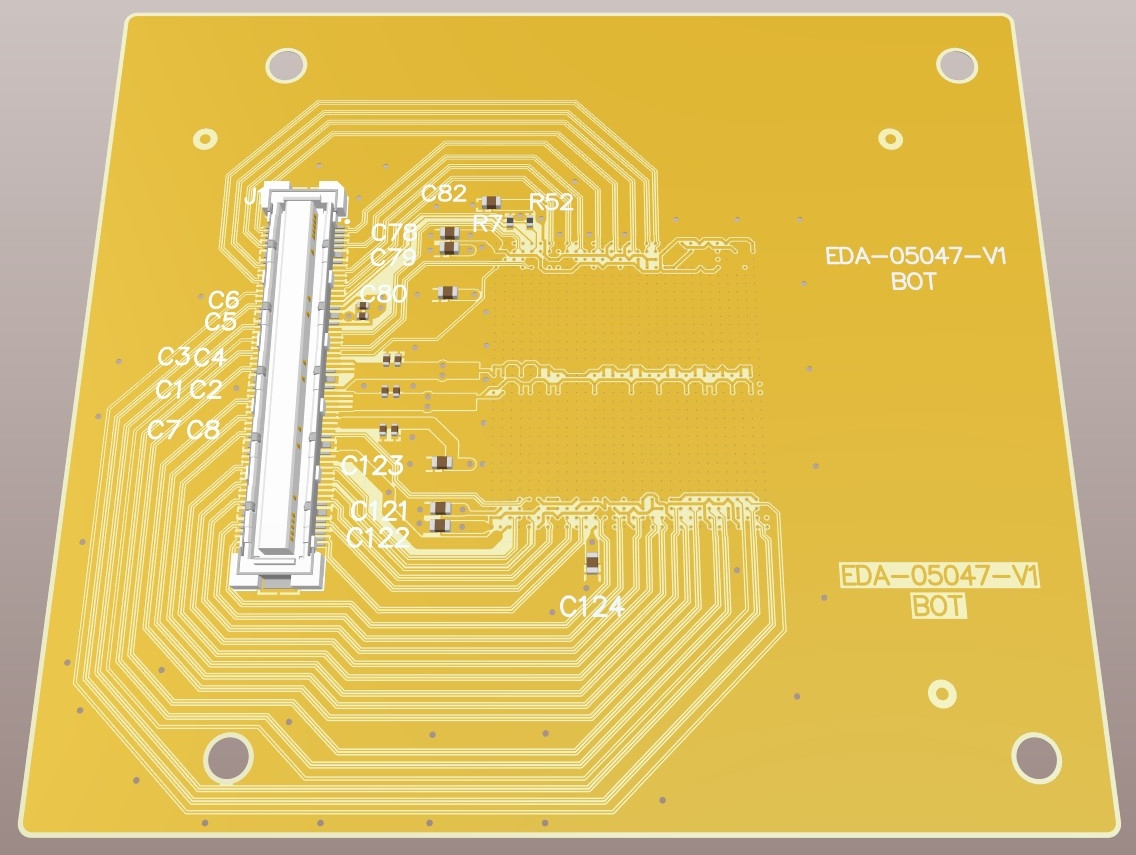}
        \caption{Flex-PCB for electrical tests}
        \label{fig:embedding-steps-1}
    \end{subfigure}
    \caption{(a) Photograph of two Timepix4 ASICs, laminated within two polyimide foils.
        (b) Drawing of the bottom side of the flex-PCB to perform the electrical tests after the embedding.}
    \label{fig:embedding-steps}
\end{figure}

Before the actual detector production, mechanical tests have been performed, showing the general feasibility of laminating a thinned\footnote{By default the Timepix4 is around $\SI{700}{\micro m}$ thick. To use the TSVs, it needs to be thinned to around $\SI{130}{\micro m}$ thickness.} and TSV-processed Timepix4 within two polyimide foils \cite{mpgd-proceeding} (Fig.~\ref{fig:embedding-steps-0}).
In a second step, which is currently ongoing, the Timepix4 will be embedded in a flexible polyimide PCB (Fig.~\ref{fig:embedding-steps-1}).
Thus, the electrical connectivity and the required slow control functionality can be tested.
For this, a dedicated detector/readout PCB was designed, which is currently in production together with the flex-PCB.
Following this, the first prototype will then included also the detector part by replacing the polyimide foil on top of the pixel matrix with a \textmu{}RWell.

\subsection{Signal induction simulation}

\begin{figure}[t!]
    \centering
    \begin{subfigure}{0.38\columnwidth}
        \centering
        \includegraphics[width = \columnwidth]{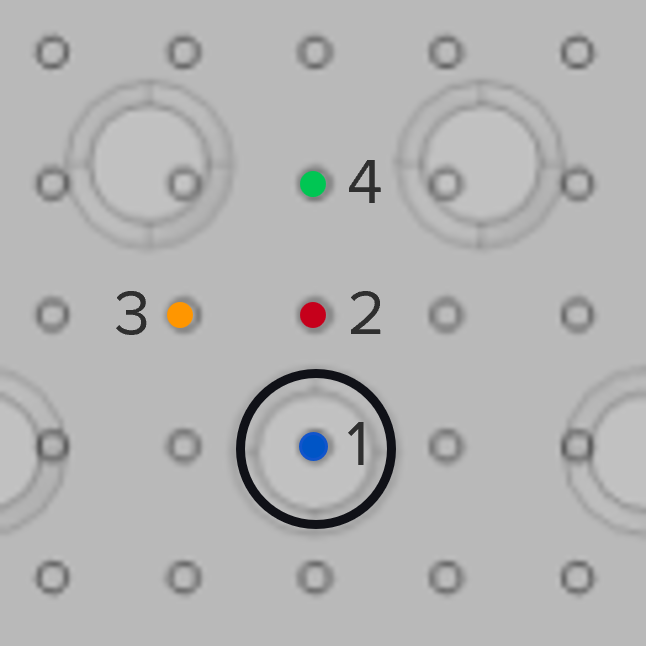}
        \caption{Geometry illustration}
    \end{subfigure}
    \hspace{0.05\columnwidth}
    \begin{subfigure}{0.40577\columnwidth}
        \centering
        \includegraphics[width = \columnwidth]{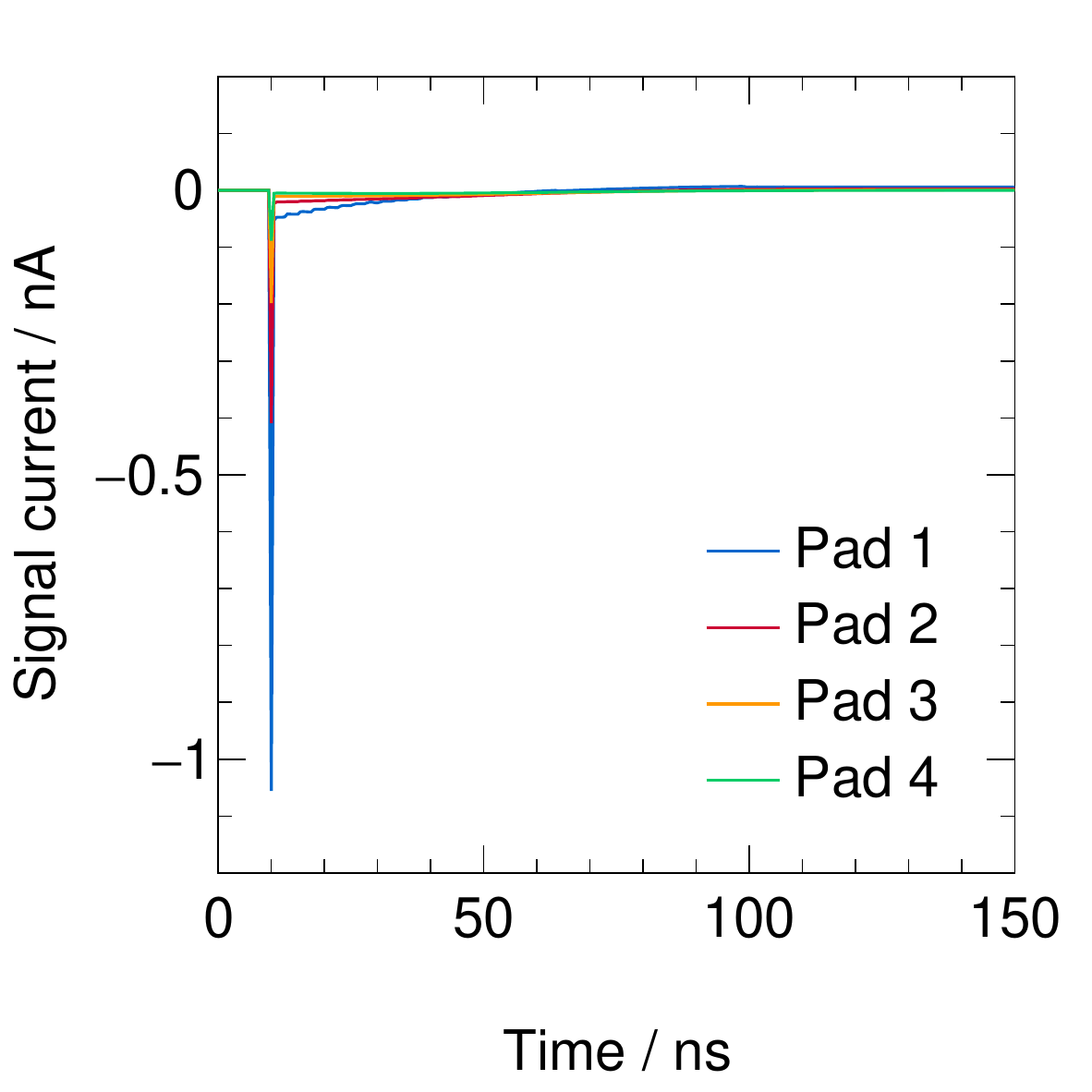}
        \caption{Induced signal current}
    \end{subfigure}
    \caption{(a) Top view of the geometry used in the signal induction simulation.
        The circle indicate the \textmu{}RWell hole in which the avalanche takes place.
        (b) Simulated signal current, induced in different readout pads.}
    \label{fig:simulation}
\end{figure}

Given the Timpeix4's square pixel grid with $\SI{55}{\micro m}$ pitch and the hexagonal hole pattern with $\SI{140}{\micro m}$ pitch of the \textmu{}RWell, it was investigated through Garfield++ simulations \cite{garfieldpp}, how these different geometries affect the signal induction (Fig.~\ref{fig:simulation}).
In one of the \textmu{}RWell holes, an avalanche was simulated.
Then, the signal current in various pixel pads, depending on the distance to the hole was calculated.
The signal current reduces significantly when the distance between the avalanche and the pixel pad is larger than the pixel pitch.
In reality, however, due to the diffusion in gaseous detectors, the charge will be spread over several neighbouring holes.
In addition, due to the higher granularity of the Timepix4 with respect to the \textmu{}RWell, it is expected that a sufficient amount of induced charge can be acquired to reconstruct the events.

\section{GEMPix4}
\label{sec:gempix}

In parallel to the embedding developments, activities on reading out a Gas Electron Multiplier (GEM) detector \cite{gem} with the Timepix4 are carried out.
This is a further development of the GEMPix technology \cite{gempix}.
These activities enables to demonstrate the capabilities of MPGDs with Timepix4 readout. 

The Nikhef carrier board, which is part of the SPIDR4, serves as a support for the detector housing with gas volume and amplification structure \cite{mpgd-proceeding}.
The triple-GEM stack inside the housing resembles the COMPASS configuration \cite{compass-gem}, with a $\SI{3}{mm}$ wide drift gap and $\SI{2}{mm}$ wide transfer and induction gaps.
It is filled with a mixture of Ar/CO\textsubscript{2} (70/30).

\subsection{Geometrical event signatures}
\begin{figure}[t]
    \centering
    \begin{subfigure}{0.45\columnwidth}
        \centering
        \includegraphics[width = \columnwidth]{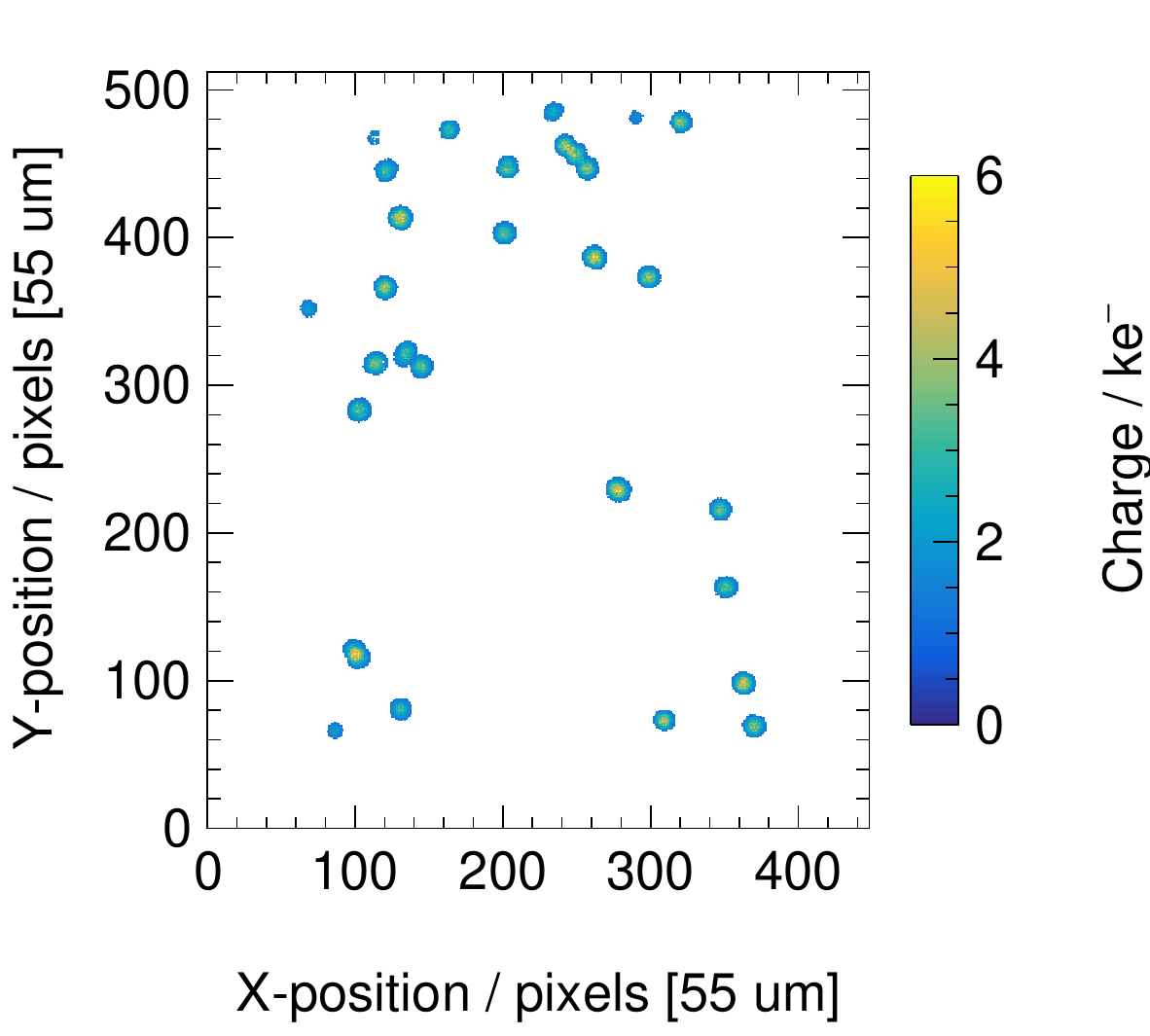}
        \caption{X-ray interactions}
        \label{fig:events-x-rays}
    \end{subfigure}
    \hspace{0.05\columnwidth}
    \begin{subfigure}{0.45\columnwidth}
        \centering
        \includegraphics[width = \columnwidth]{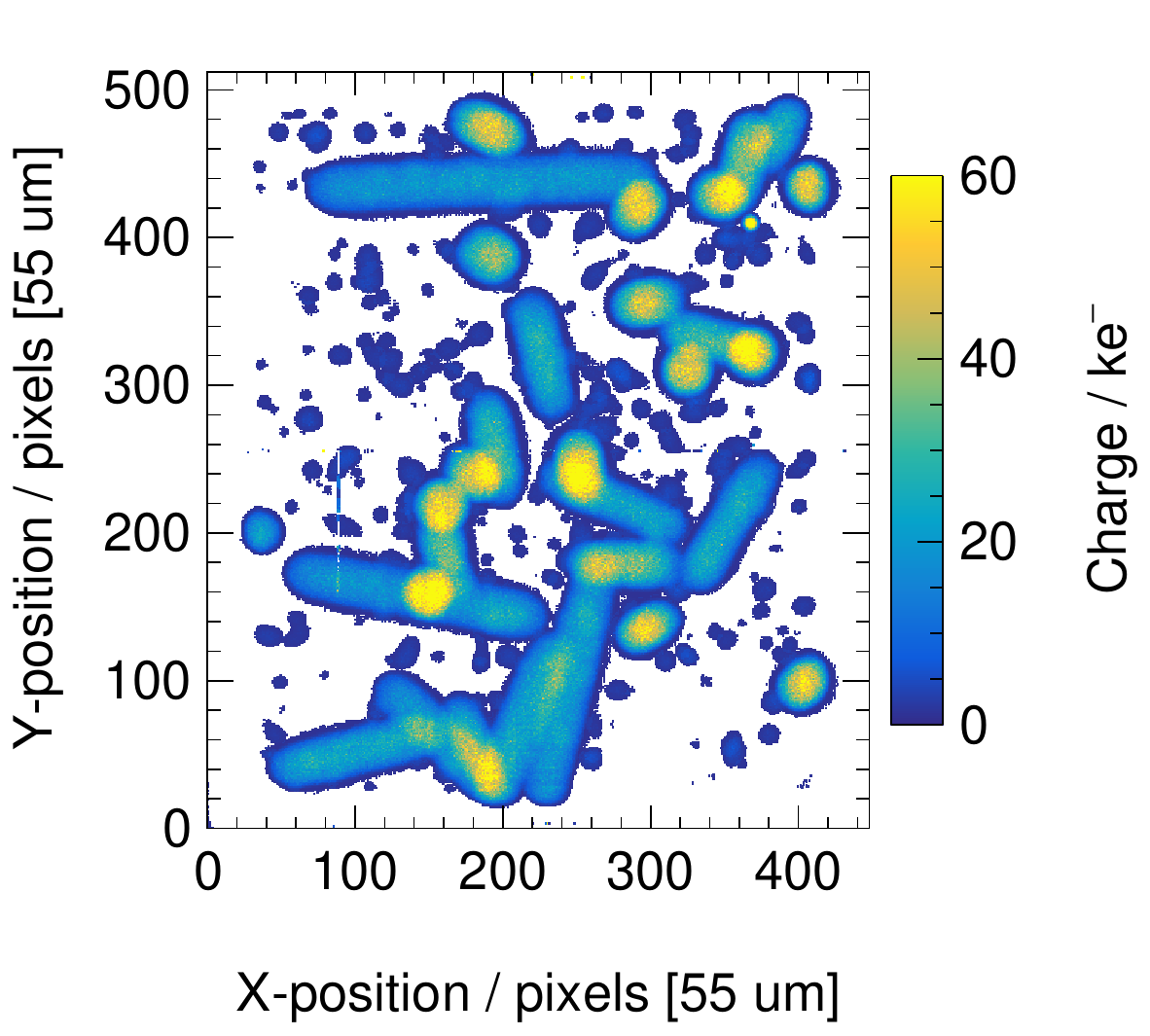}
        \caption{Cosmic ray interactions}
        \label{fig:events-cosmics}
    \end{subfigure}
    \caption{Example interactions from $\SI{6}{keV}$ X-rays (a) and cosmic rays (b) recorded with the GEMPix in a gas mixture of Ar/CO\textsubscript{2} (70/30).
        The colour scale corresponds to the charge collected in each pixel.}
    \label{fig:events}
\end{figure}

A key aspect of combining gaseous detector and pixel readout is the event identification purely based on geometrical parameters (Fig.~\ref{fig:events}).
Here, it can be seen that in Ar/CO\textsubscript{2} (70/30) soft X-rays produce circular shapes with around $\SI{1}{mm}$ diameter and $\num{e2}$ active pixels. 
Cosmic rays can result in similar shapes, which are most likely caused by Minimum Ionising Particles (MIPs) traversing the detector perpendicular to the readout plane.
However, as soon as the charged particle is traversing the detector not perpendicular, the geometrical shape becomes more asymmetric. 
In addition, various interactions from highly ionising particles can be seen resulting in long tracks or elliptical shapes with $>\num{e3}$ active pixels and much more collected charge per pixel compared to X-rays or MIPs.

\subsection{X-ray radiography}
Using both readout modes, various X-ray imaging studies have been performed, showing here only the results obtained with a line-pair phantom.
\begin{figure}[t]
    \centering
    \begin{subfigure}{0.45\columnwidth}
        \centering
        \includegraphics[width = \columnwidth]{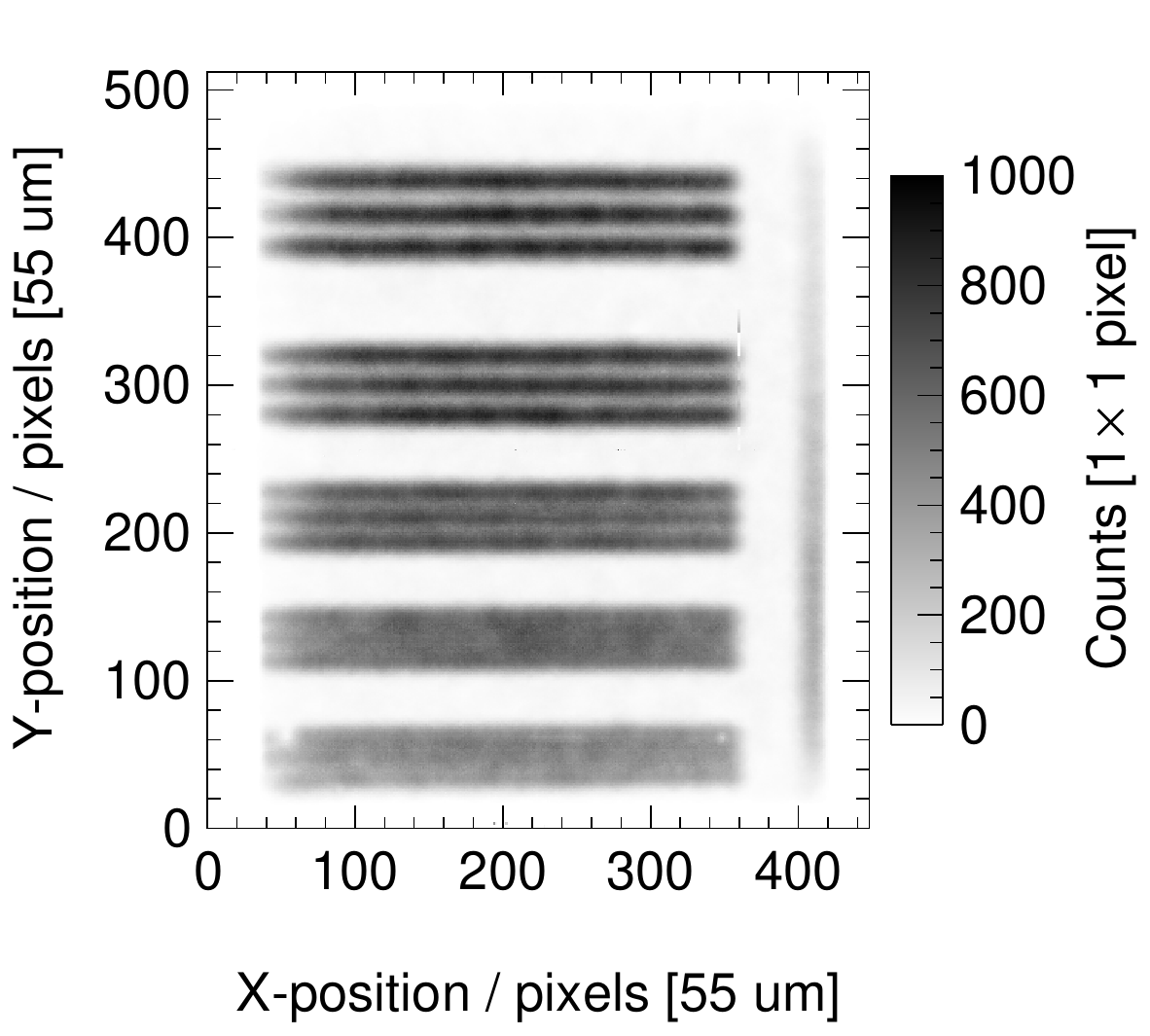}
        \caption{Integrated pixels}
        \label{fig:x-ray-imaging-frame}
    \end{subfigure}
    \hspace{0.05\columnwidth}
    \begin{subfigure}{0.45\columnwidth}
        \centering
        \includegraphics[width = \columnwidth]{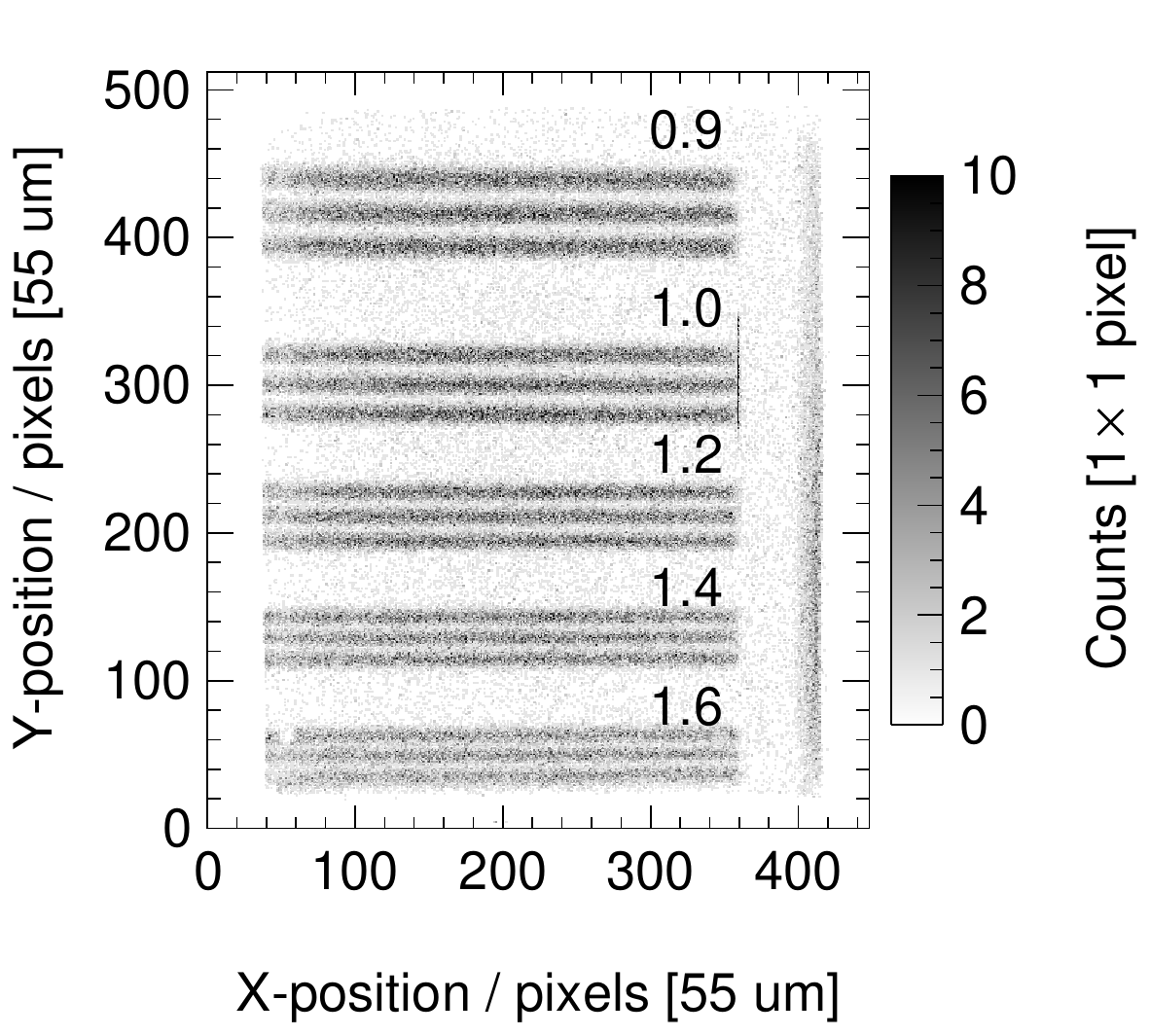}
        \caption{Interaction centres only}
        \label{fig:x-ray-imaging-data}
    \end{subfigure}
    \caption{(a) Shows an integrated image of a line-pair phantom.
    (b) Shows only the interaction points, which have been determined from each circular interaction shape, using the charge centroid method.
    The annotated values indicate the number of line pairs per millimetre (LP/mm).}
    \label{fig:x-ray-imaging}
\end{figure}
While in the frame-based readout mode, reasonable images can be obtained, it is still affected by the diffusion in the gas.
This effect can be significantly reduced when performing event reconstruction and only plotting the centre of each cluster (see the comparison between Fig.~\ref{fig:x-ray-imaging-frame} and Fig.~\ref{fig:x-ray-imaging-data}).

These measurements could additionally profit from a detector with a single amplification stage, like the \textmu{}RWell from the embedding approach.
Due to the missing transfer and induction gaps, only the drift region remains, i.e. the maximum drift distance would be reduced from $\SI{9}{mm}$ to $\SI{3}{mm}$, leading also to small diffusion effects.

\subsection{Energy resolved studies}
To demonstrate the ability of detecting low energetic X-rays, studies to resolve $\SI{1.5}{keV}$ aluminium fluorescence X-ray have been conducted.
To ensure that these photons can be detected within the drift region, the detector's $\SI{100}{\micro m}$ thick polyimide entrance window was replaced with a $\SI{2}{\micro m}$ thin Mylar foil, and the polyimide-copper cathode was replaced with a steel mesh. 

\begin{figure}[t!]
    \centering
    \includegraphics[width = \columnwidth]{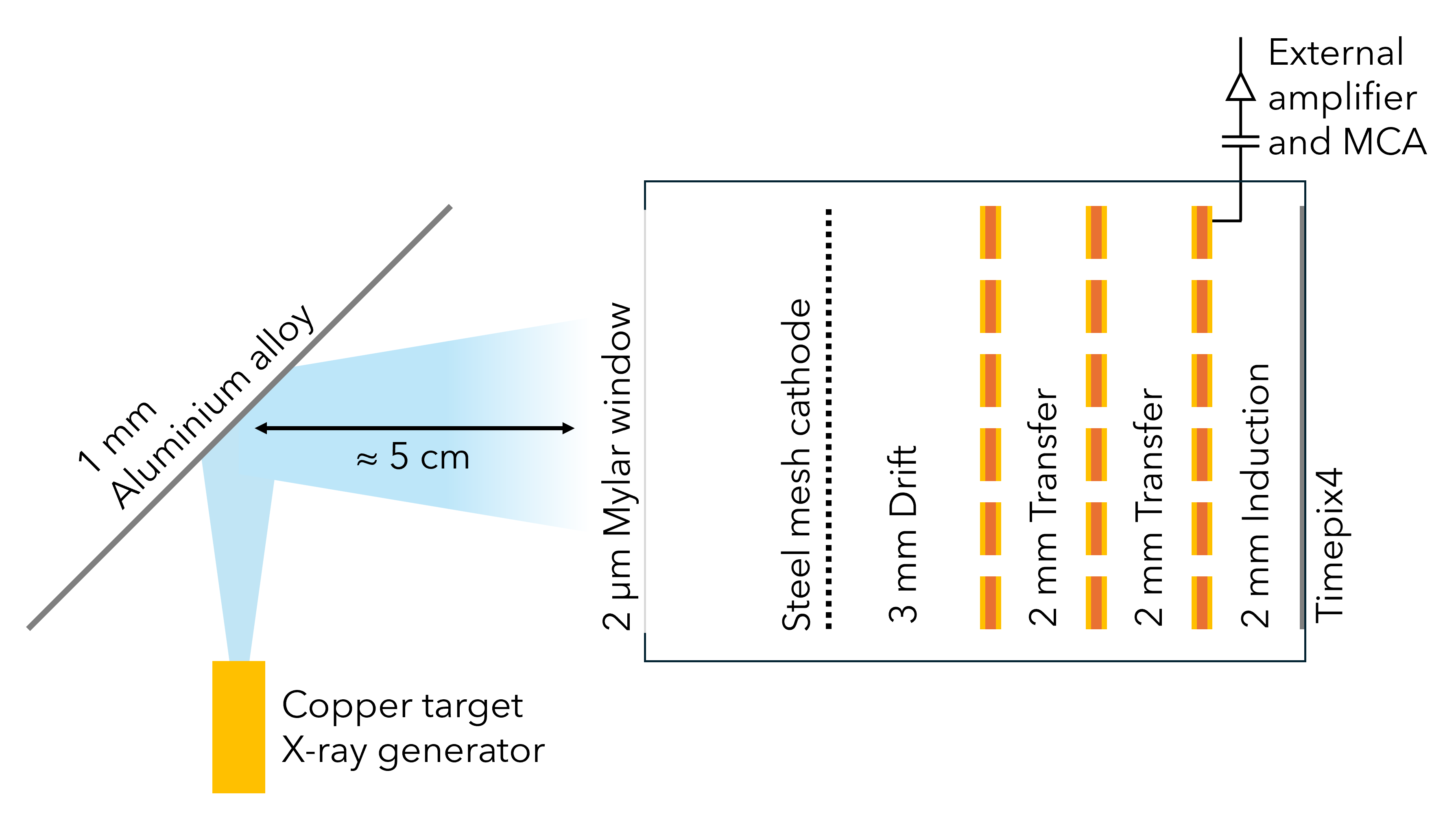}
    \caption{Schematic illustration of the experimental setup used for the X-ray fluorescence measurements (not drawn to scale).}
    \label{fig:x-ray-setup}
\end{figure}

\begin{figure}[t]
    \centering
    \begin{subfigure}{0.40577\columnwidth}
        \centering
        \includegraphics[width = \columnwidth]{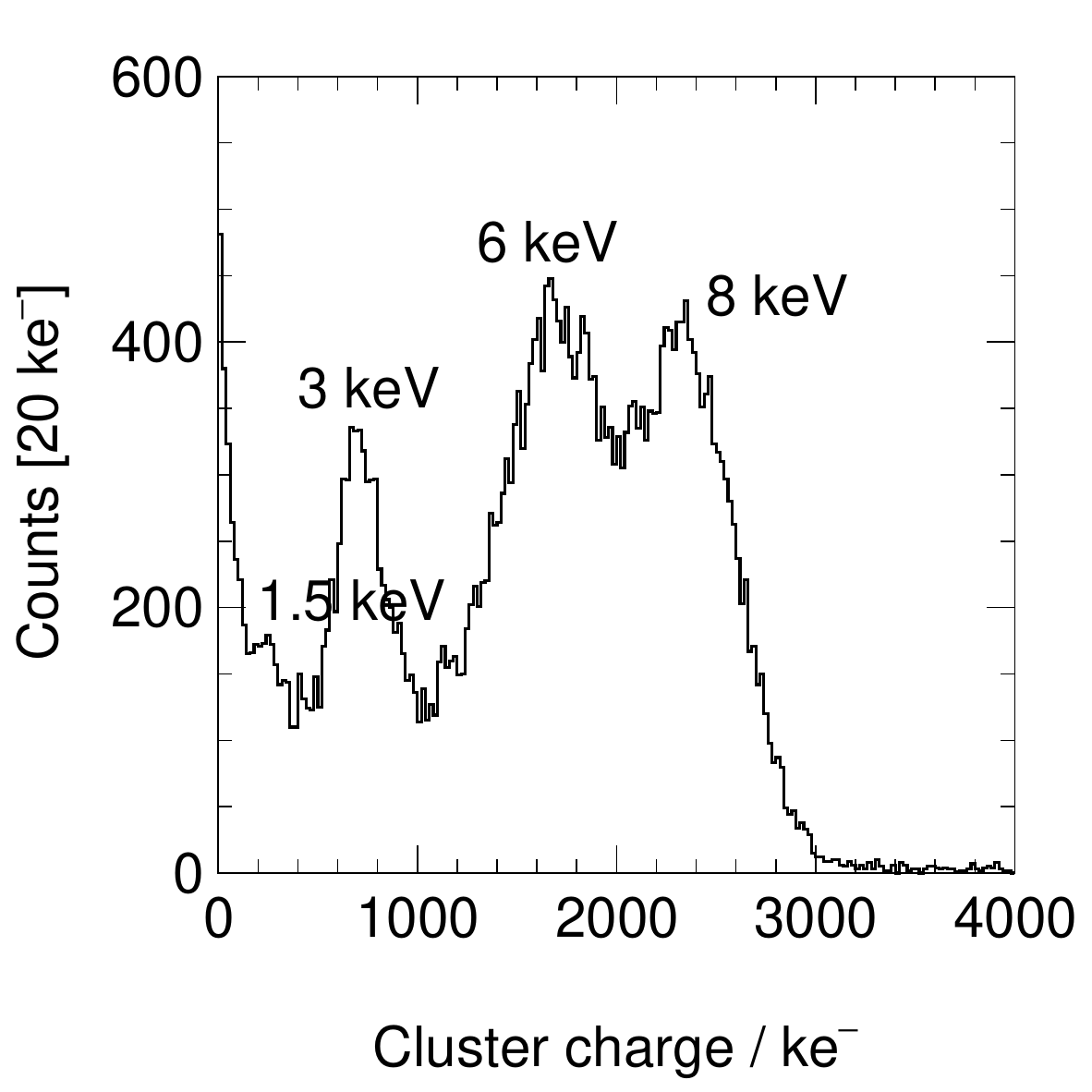}
        \caption{Timepix4 readout}
        \label{fig:energy-resolved-external-gempix}
    \end{subfigure}
    \hspace{0.05\columnwidth}
    \begin{subfigure}{0.40577\columnwidth}
        \centering
        \includegraphics[width = \columnwidth]{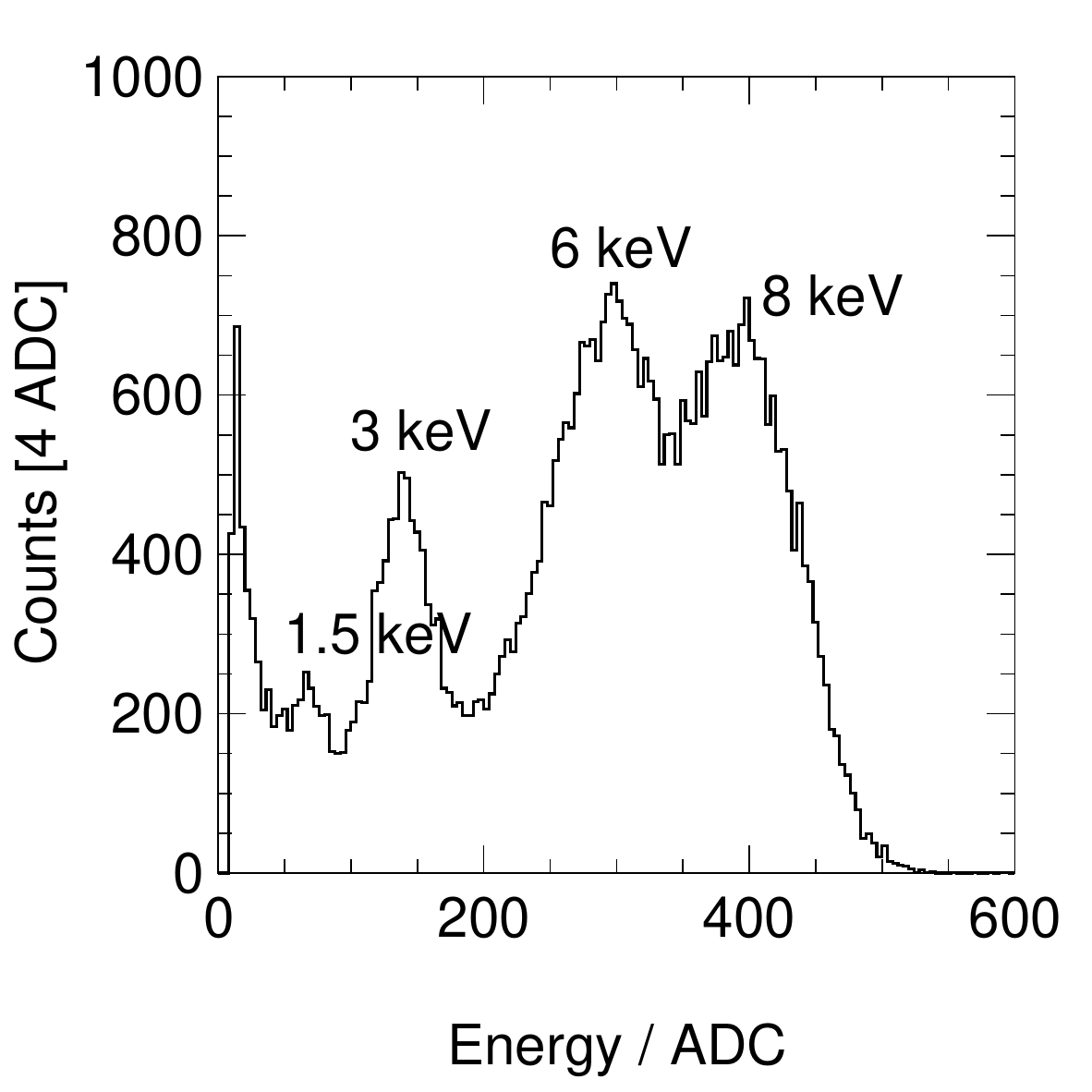}
        \caption{External MCA}
        \label{fig:energy-resolved-external-mca}
    \end{subfigure}
    \caption{Fluorescence spectra measured with (a) the Timepix4 and (b) the external MCA connected to the bottom of the third GEM.}
    \label{fig:energy-resolved-external}
\end{figure}

For the measurements, an aluminium alloy plate was irradiated with X-rays generated by a copper target X-ray tube (see a sketch of the setup in Fig.~\ref{fig:x-ray-setup}).
The resulting fluorescence spectrum (Fig.~\ref{fig:energy-resolved-external-gempix}) 
was acquired with the GEMPix4.
Simultaneously, the spectrum has been acquired from the bottom electrode of the third GEM (Fig.~\ref{fig:energy-resolved-external-mca}), using a discrete single-channel pre-amplifier and shaper electronics, connected to an external Multi-Channel Analyser (MCA).
This serves as reference for the detector response.

In both spectra, various X-ray energies can be resolved.
The line at $\SI{8}{keV}$ corresponds to deflected copper X-rays from the copper target X-ray tube.
The fluorescence line at $\SI{6}{keV}$ originates from the presence of manganese and iron in the alloy.
The line at $\SI{3}{keV}$ originates from argon fluorescence in the gas volume between the entrance window and mesh cathode.
It also corresponds to the so-called `argon escape peak' of the $\SI{6}{keV}$ manganese X-rays.
The line at $\SI{1.5}{keV}$ corresponds to the aluminium fluorescence X-rays.
Due to their strong absorption in the $\sim\SI{5}{cm}$ of air between the aluminium plate and detector, this line is much less pronounced than the other lines.

It should be mentioned that the ratios between the different energy lines are correct for the spectrum acquired with the MCA, but they are not correct for the spectrum acquired with the Timepix4.
Requiring further investigation, possible explanations for this behaviour are threshold effects on the Timepix4, as well as an insufficient charge calibration procedure.

\begin{figure}[t]
    \centering
    \begin{subfigure}{0.40577\columnwidth}
        \centering
        \includegraphics[width = \columnwidth]{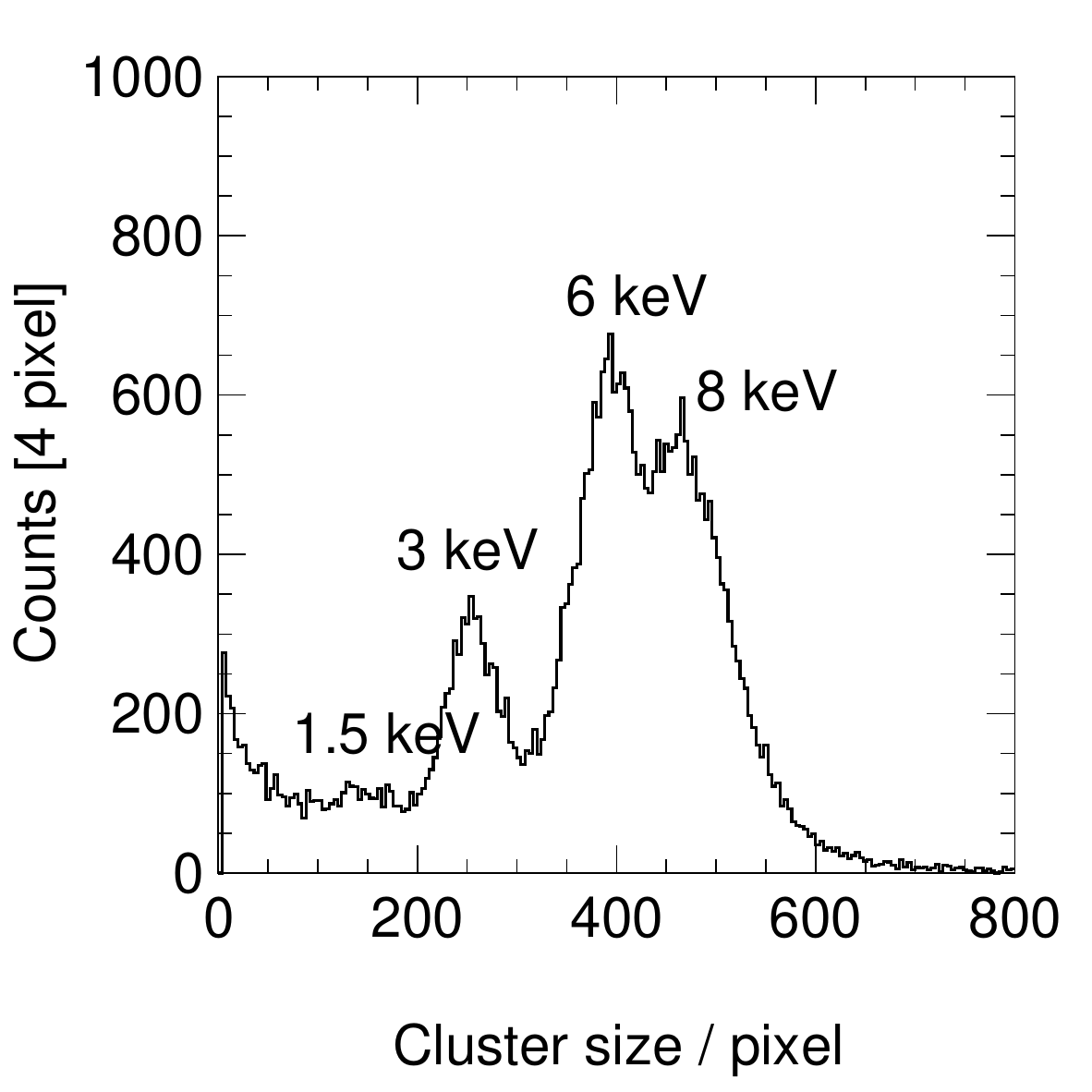}
        \caption{Cluster size}
        \label{fig:energy-resolved-pixel-spectrum}
    \end{subfigure}
    \hspace{0.05\columnwidth}
    \begin{subfigure}{0.486925\columnwidth}
        \centering
        \includegraphics[width = \columnwidth]{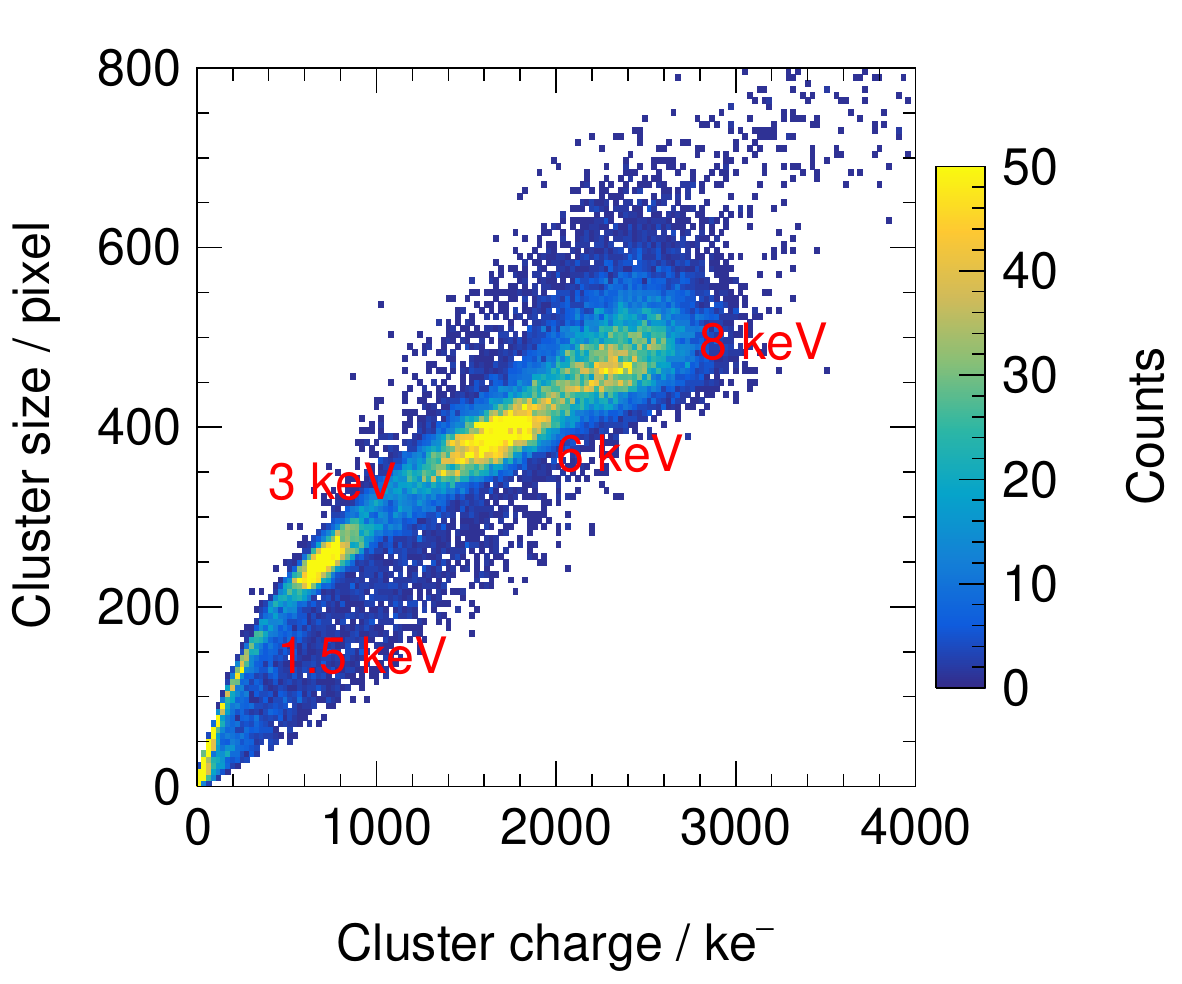}
        \caption{Cluster charge-size correlation}
        \label{fig:energy-resolved-pixel-correlation}
    \end{subfigure}
    \caption{(a) Cluster size spectrum for the fluorescence interactions.
        (b) Correlation between cluster size and cluster charge.}
    \label{fig:energy-resolved-pixel}
\end{figure}

In addition to the charge information, also the cluster size was studied, i.e. the number of active pixels per interaction (Fig.~\ref{fig:energy-resolved-pixel-spectrum}).
The histogram shows also an energy dependence.
Thus, the correlation between total charge and number of active pixels per photon was studied (Fig.~\ref{fig:energy-resolved-pixel-correlation}), showing a non-linear behaviour.
This requires further investigation, possibly with a lower-density gas mixture which reduces multiple scattering.

\subsection{Time resolved studies}
\begin{figure}[t]
    \centering
    \includegraphics[width = 0.45\columnwidth]{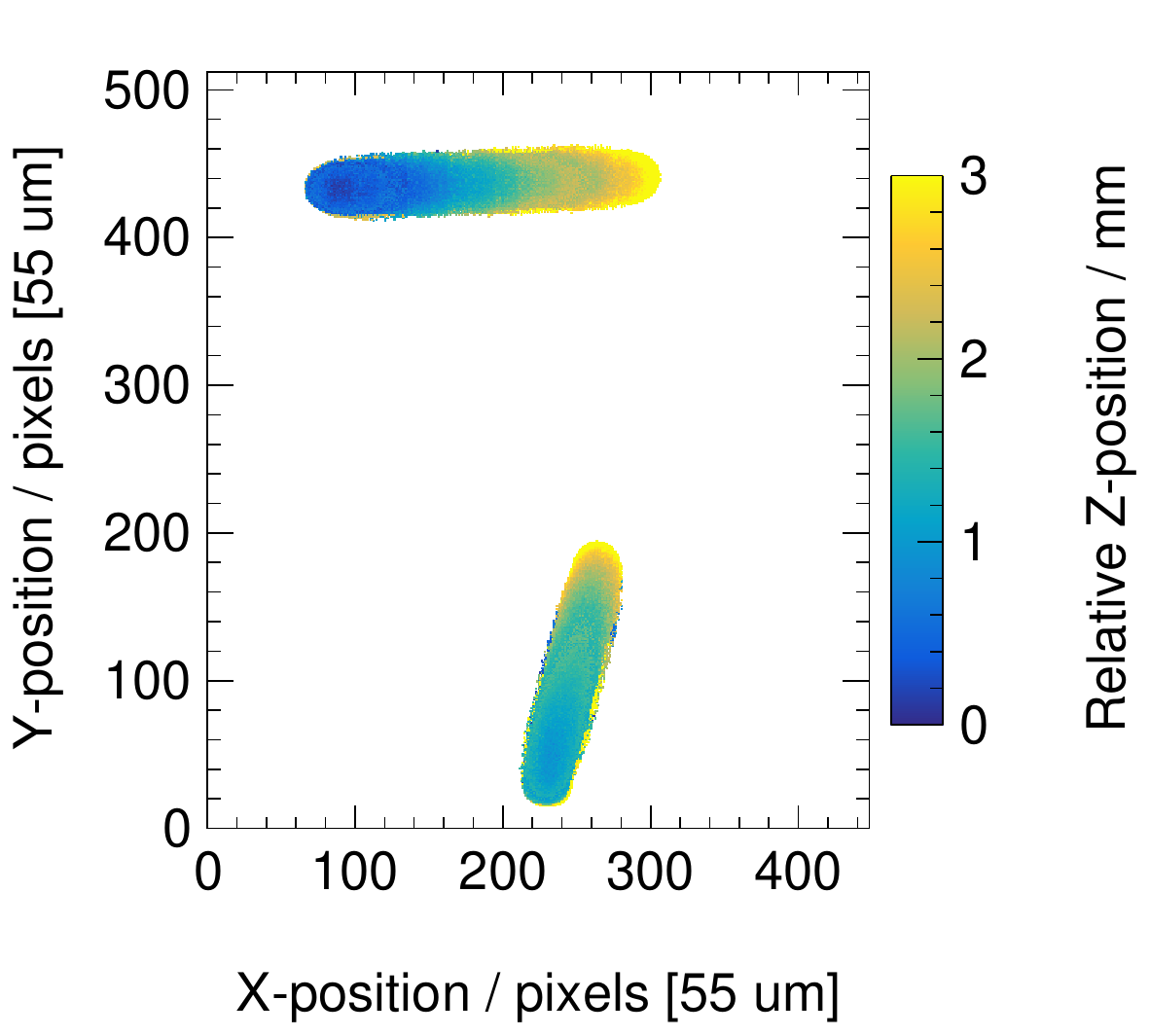}
    \caption{Interaction depth of the highly ionising particles along their trajectory through the drift region within the GEMPix4 detector.}
    \label{fig:tpc}
\end{figure}

To investigate the timing behaviour of the GEMPix4 detector and profit from the time resolution of the Timepix4, two tracks from the highly ionising particles (Fig.~\ref{fig:events-cosmics}) have been selected.
For both tracks, the TOA values were converted into depth information, using the electron drift velocity.
The result is shown in Fig.~\ref{fig:tpc}.
It can be seen that both tracks have an interaction depth from $\num{0}$ to $\SI{3}{mm}$, as expected from the detector geometry.


\section{Conclusion and outlook}

In this paper, some first results of high-granularity charge readout of MPGDs using the Timepix4 have been presented.
In combination with a triple-GEM detector, X-ray imaging, energy-resolved studies (measuring X-ray photons with energies below $\SI{2}{keV}$) and time-resolved studies (profiting from the $\SI{200}{ps}$ time bins of the Timepix4), as well as for particle identification, based on geometrical parameters have been performed.
In parallel, the embedding of the Timepix4 into a micro-pattern amplification stage, specifically a micro-resistive Well, using standard PCB technologies is investigated.
The first mechanical tests have been successfully performed at CERN's MPT workshop.
Furthermore, it is planned to investigate the feasibility of integrating other pixel ASICs into MPGDs, possibly also wafer-scale ASICs.

\section*{Acknowledgements}
This work has been supported by the CERN EP R\&D Strategic Programme on Technologies for Future Experiments (\url{https://ep-rnd.web.cern.ch/}).

\bibliography{references}

\end{document}